\documentclass[pdflatex,sn-mathphys]{sn-jnl}

\jyear{2021}%

\theoremstyle{thmstyleone}%
\theoremstyle{thmstyletwo}%

\theoremstyle{thmstylethree}%

\begin{document}

\title[  ]{Designing Topological Defect Lines Protected by Gauge-dependent Symmetry Indicators}


\author[1]{\fnm{Erda} \sur{Wen}}\email{ewen@ucsd.edu}
\equalcont{These authors contributed equally to this work.}

\author[2]{\fnm{Dia'aaldin J.} \sur{Bisharat}}\email{iiauthor@gmail.com}
\equalcont{These authors contributed equally to this work.}

\author[1]{\fnm{Robert J.} \sur{ Davis}}\email{rjdavis@ucsd.edu}
\equalcont{These authors contributed equally to this work.}

\author[1]{\fnm{Xiaozhen} \sur{Yang}}\email{xiy003@ucsd.edu}

\author* [1]{\fnm{Daniel F.} \sur{Sievenpiper}}\email{dsievenpiper@ucsd.edu}

\affil*[1]{\orgdiv{Electrical and Computer Engineering}, \orgname{University of California, San Diego}, \orgaddress{\street{9500 Gilman Drive \#0407}, \city{La Jolla}, \postcode{92093}, \state{CA}, \country{USA}}}

\affil[2]{\orgdiv{Advanced Science Research Center}, \orgname{The City University of New York}, \orgaddress{\street{85 St Nicholas Terrace}, \city{New York}, \postcode{10031}, \state{New York}, \country{USA}}}


\abstract{Symmetry indicators are a modern tool for characterizing topological phases that require only minimal computational expense but provide an elegant means of designing practical devices. This paper demonstrates how a rotational symmetry indicator can be used to construct and characterize a topologically robust waveguide, which is then verified experimentally on a printed circuit board (PCB) platform. The design takes advantage of the real-space gauge-dependency of the symmetry indicators and adopts a $C_6$ lattice with simple shifts, forming a defect line supporting topological edge modes. It is shown that the modes can realize the same features as previous topological waveguides, but in addition possesses a greater degree of reconfigurability and the unique ability to form a one-way termination. Moreover, the design illustrates the critical role real space information plays in determining the topological properties of photonic crystals, enabling a wider range of possible realizations.}




\maketitle

\section{Introduction}\label{sec1}

Bringing the concept of ``topological phases'' in condensed matter physics to photonics\cite{Ozawa2019,Kim2020,Davis2021} provides a new perspective for light wave manipulation, and inspires a vast number of unprecedented optical device designs with intriguing properties, including waveguides with enhanced robustness against imperfections and high-Q nanocavities\cite{Wang2009,Khanikaev2013,Ota2019}. These designs appear in different forms depending on the underlying symmetries\cite{Varjas2017}, with various designs being characterized by topological invariants like the Chern number\cite{Wang2009}, spin/valley-Chern numbers\cite{Nalitov2015,Ma2016,Dong2017,Wu2017,Orazbayev2018,Noh2018,Bisharat2019,Xue2021}, and more recently by quantized Zak phase\cite{Liu2017,Liu2018,Ota2019} and Weyl points\cite{Chen2016,Wang2019}. A nonzero value of these invariants naturally gives rise to topologically protected edge states or higher-order states via the bulk-edge or edge-corner correspondence \cite{Hatsugai1993,Fu2007,Liu2021}. Despite the fact that these designs are well explained and some are elegant representations of their electronic counterparts, i.e. integer/valley/spin quantum Hall effects, recent works demonstrate various anomalous topological designs that appear beyond these principles, such as anti-phase boundaries\cite{Kong2020} and local-valley interfaces [Bisharat2021], with materials that are regraded as trivial according to the metrics aforementioned. As these metrics struggle to give a generalized explanation for such platforms, they are consequently inadequate in predicting new topological crystalline arrangement in a general manner as well.

On the other hand, similar difficulty also arises in identifying and predicting new topological crystalline insulators (TCIs)\cite{Fu2011} with these theories that require numerical calculation of the full wave-function across the whole momentum space, which can be computationally expensive given the TCIs' richness of symmetry. This leads to more recent studies that propose to address the issue with the symmetry indicator approach\cite{Fang2012,Taherinejad2014,Kruthoff2017,Benalcazar2019,Po2020}, which at its core, focuses on determining the topology of the bandstructure via the symmetries of eigenstaes found at high symmetry points (HSPs), rather than across the full Brillouin zone. Moreover, such tools have the power to explain systems that have otherwise vanishing topological invariants, as features like inversion or rotaional crystaline symmeties may be the cause instead. The reduced complexity in calculation allows it to reveal a more complete catalogue of topological materials\cite{Tang2019,Vergnoiory2019,Zhang2019}, and moreover, an efficient and broadened means to construct crystals with non-trivial topology.

Remarkably, an atypical property of the symmetry indicator is its dependency on the selection of the lattice boundary: different unit-cell representations of the same lattice may result in different topology. This manifests a fundamental aspect of TCIs in that the real-space information contained in the basis selection plays a critical role in determining the topological properties of them. Earlier studies on this can be found in \cite{Ota2019,Ni2019} where it is argued that two choices of unit cell that result in identical infinite lattices but are inequivalent when a finite boundary is added can posses differing bulk topologies based on the Zak phase and generalized chiral symmetry, respectively. A more generalized statement in \cite{Davis2022} further clarifies the importance of the localized effect at the boundary of TCIs with the notion of rotational symmetry indicators, and the possibility to engineer non-trivial indicators by merely shifting an otherwise trivial unit cell.

The objective of this paper is to demonstrate that this new principle is capable of creating very practical photonic designs, starting from a general tight-binding model and proceeding to a printed circuit board (PCB) design that supports topological edge state for surface waves. In particular, the lattice adopted on each side of the waveguide discussed in this work are identical trivial $C_6$-symmetric lattice, but by introducing a certain real-space offset between them, two inequivalent $C_3$-symmetric structures with non-trivial topology are induced. We will also show how this platform can be employed for unique arrangements such as a topological waveguide with a termination on one end, which cannot be obtained with conventional topological waveguides.

\section{Results}\label{sec2}

\subsection{$C_6$ lattice tight-binding model}
\begin{figure}[ht]
\centering
\includegraphics[width=12cm]{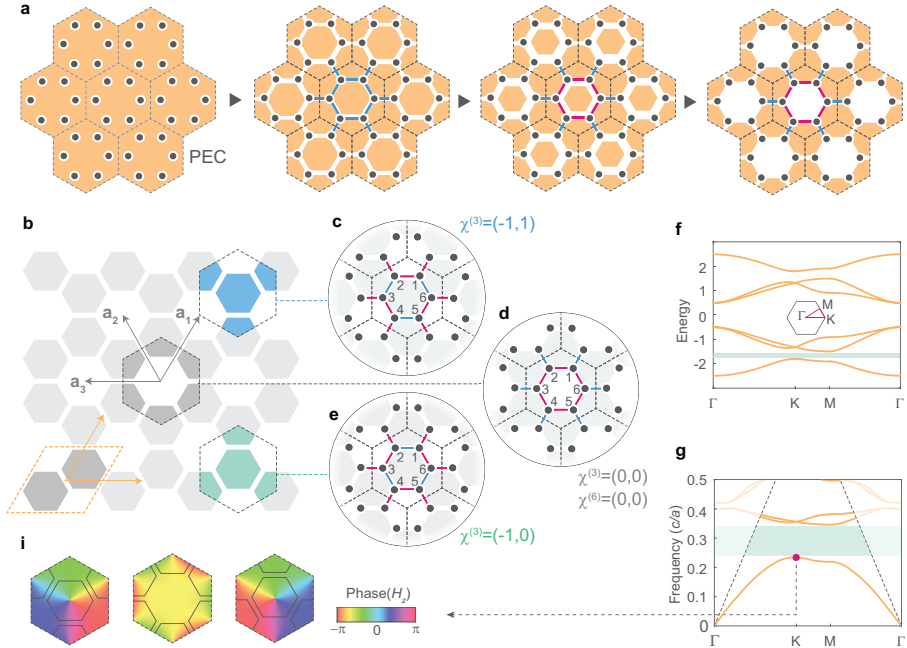}
\caption{The lattice designs. \textbf{a} The step-by-step design process of the metallic geometry resembling a TB Model with hexagonal lattice consisting six sites in each unit. Blue line and red line represent intra-unit/inter-unit coupling of different intensity. \textbf{b} Different unit-cell boundary selection leads to different topology: \textbf{d} Topological trivial unit with strong intra-unit coupling and weak inter-unit coupling; \textbf{c} and \textbf{e} Topological non-trivial unit with strong inter-unit coupling an hybrid intra-unit coupling. \textbf{f} Band diagram of the tight-binding model with $\gamma=1$ and $\lambda=0.5$. \textbf{g} Dispersion relationship of the metallic geometry on the substrate, the gap size $g_2=0.04a$. The dotted line represents the light cone. The bandgap being inspected is highlighted in \textbf{f} and \textbf{g}. \textbf{i} Phase plot of transverse H-field at K point for the units in \textbf{c}-\textbf{e}.}
\label{fig_unit}
\end{figure}

The tight-binding model to be examined and implemented in this work is a 2-D hexagonal lattice consisting of six sites in each unit with both intra-cell coupling and inter-cell coupling as shown in Fig. \ref{fig_unit}d. The (spinless) Hamiltonian of the system is given by:
\begin{equation}
H^{(6)}(\mathbf{k}) = 
\begin{pmatrix}
0 & \gamma & 0 & \lambda e^{i\mathbf{a_1\cdot k}} & 0 & \gamma\\
\gamma & 0 & \gamma &  0 &\lambda e^{i\mathbf{a_2\cdot k}} & 0\\
0 & \gamma & 0 & \gamma & 0 &\lambda e^{i\mathbf{a_3\cdot k}} \\
\lambda e^{-i\mathbf{a_1\cdot k}} & 0 & \gamma & 0 & \lambda & 0\\
0 & \lambda e^{-i\mathbf{a_2\cdot k}} & 0 & \gamma & 0 & \gamma\\
\gamma & 0 & \lambda e^{-i\mathbf{a_3\cdot k}} & 0 & \gamma & 0\\
\end{pmatrix},
\label{eq_h6}
\end{equation}
where $\gamma$ and and $\lambda$ represent the intra-cell coupling and the inter-cell coupling respectively, and $\mathbf{a_1}$, $\mathbf{a_2}$, $\mathbf{a_3}$ are illustrated in Fig. \ref{fig_unit}b. A zero on-site energy is assumed here which does not affect the generality of the conclusion. Here we choose the hopping ratio $\mid\gamma\mid>\mid\lambda\mid$ so that a band gap exists between the two lower bands, as in Fig. \ref{fig_unit}f.  This Hamiltonian has both time reversal symmetry and inversion symmetry (from the $C_6$ rotational invariance), and therefore has identically zero Berry curvature \cite{Liu2017}. Correspondingly all gauge-invariant topological indices (Chern/valley Chern number) are also zero. Note that as we are investigating the properties of the lowest bandgap there is no two-fold degeneracy that can result in a pseudospin degree of freedom as was described earlier \cite{Wu2015,Liu2021}. Now we may consider shifting the unit-cell boundary position as illustrated in Fig. \ref{fig_unit}c or Fig. \ref{fig_unit}e, assuming the lattice is infinite. The unit cell then becomes $C_3$-symmetric with strong inter-cell coupling and a hybrid intra-cell coupling. The Hamiltonian therefore appears in a different form, taking the former case as an example:
\begin{equation}
H^{(3)}(\mathbf{k}) = 
\begin{pmatrix}
0 & \gamma & 0 & \gamma e^{i\mathbf{a_1\cdot k}} & 0 & \lambda\\
\gamma & 0 & \lambda &  0 &\gamma e^{i\mathbf{a_2\cdot k}} & 0\\
0 & \lambda & 0 & \gamma & 0 &\gamma e^{i\mathbf{a_3\cdot k}} \\
\gamma e^{-i\mathbf{a_1\cdot k}} & 0 & \gamma & 0 & \lambda & 0\\
0 & \gamma e^{-i\mathbf{a_2\cdot k}} & 0 & \lambda & 0 & \gamma\\
\lambda & 0 & \gamma e^{-i\mathbf{a_3\cdot k}} & 0 & \gamma & 0\\
\end{pmatrix}.
\label{eq_h3}
\end{equation}

Under periodic boundaries, both Hamiltonians yield the same band structure. However, we can distinguish between the two systems by considering the symmetry indicators. The symmetry indicators of the $C_3$ and $C_6$ structures are given by \cite{Bradlyn2017,Po2017,Benalcazar2019,Li2020} $\chi^{(3)} = \left(\left[K_1^{(3)}\right],\left[K_2^{(3)}\right]\right)$ and $\chi^{(6)} = \left(\left[M_1^{(2)}\right],\left[K_1^{(3)}\right]\right)$, in which$\left[\Pi_p^{(n)}\right] \equiv \#\Pi_p^{(n)}-\#\Gamma_p^{(n)}$, $\#\Pi_p^{(n)}$ being the number of bands below the bandgap at high-symmetry points, $K$ or $M$ point as in these cases, with the rotation eigenvalue $\Pi_p^{(n)}=e^{2\pi i (p-1)/n}, (p = 1,\dots,n)$. Under this notation, the symmetry indicator for the $H^{(3)}$ can be calculated to be $\chi^{(3)} = (-1,+1)$, indicating a non-trivial topology. Similarly, by first applying a $C_2$ rotation of the basis in real space, an opposite symmetry indicator $\chi^{(3)} = (-1,0)$ can be obtained for case in Fig. \ref{fig_unit}e. For $H^{(6)}$, the symmetry indicator is found to be $\chi^{(6)} = (0,0)$. In order to evaluate its interaction with the shifted unit-cell, we may also ``downgrade'' it to $C_3$ symmetry and still find $\chi^{(3)} = (0,0)$. Both indicate a trivial topology of the unshifted cell. It will become clear in the following sections that this intriguing difference in topology corresponds to a localized effect at different interface in a finite lattice.

\subsection{TB model realization with metallic surface}
Since the practicality of the design is one of the main concerns of this work, we forgo the 2-D pure dielectric structure proposed in previous studies\cite{Wu2015,Yang2018,Liu2021}, which require certain boundaries in the $z$-axis and are usually bulky. Instead, we demonstrate a new approach to realize this with metallic surfaces that can be printed on a PCB, the design process of which is illustrated in Fig. \ref{fig_unit}a. Consider a hexagonal unit cell with six separated holes and an equal spacing of $d=a/3$, as the fist inset illustrates. Each opening, being a small cavity where the field is confined, resembles an independent atomic site, with little hopping between them. We may then introduce the interaction between these sites by creating channels that connect them, allowing energy coupling,  the intensity of which can be controlled by the width of the gap, as shown in the third inset. A larger separation corresponds to a stronger coupling, as shown in Supplementary Note 1, and the topological phase transition indeed happens when intra-unit gap size $g_1$ equals to inter-unit gap size $g_2$, where the bandgap closes at the $K$ point to form a degeneracy.

An extreme case of the strong intra-unit coupling and weak inter-cell coupling is when the center patch in each unit-cell vanishes so that $g_1$ is maximized, as shown in the fourth inset. This setup will be utilized in the following part for its simplicity in geometry, but we would like to point out that any general selection of $g_1$ and $g_2$ satisfying $g_1>g_2$, or any other methods that differentiate the coupling intensity, should demonstrate the same behavior.

The structure is printed on a $0.125a$-thick substrate with a high dielectric constant of 10.2 (See Methods, PCB fabrication). This choice was made to bring the frequency range of the bandgap down below the light-cone, so that surface mode can be properly excited and supported, as shown in Fig. \ref{fig_unit}g. This also has the benefit  over the earlier shift-lattice model\cite{Kong2020} that operates near the $\Gamma$ point, where inducing surface wave becomes impossible. Nevertheless, the adoption of the substrate is not essential for observing topological edge states. Under this configuration, the modes exist as $TE_z$ surfaces waves that are confined in the gaps. The horizontal $E$ field causes a potential difference between each hexagonal patch, which permits a simple means to excite the surface waves via RF sources across the patches. 

The topological symmetry indicator can also be verified here with the numerical solution of the eigenmode for the structure straightforwardly by looking at the phase of the horizontal magnetic field at the $K$ point, as in Fig. \ref{fig_unit}i: for the topologically non-trivial cases, we observe a phase progression around the center that exhibits a rotational eigenvalue $K_2^{(3)}=e^{i\frac{\pi}{3}}$, while for the topologically trivial case, the eigenvalue is $K_1^{(3)}=1$. For the $C_2$-rotated structure we also observe a reversal of the phase winding, which gives $K_3^{(3)}=e^{i\frac{2\pi}{3}}$. This observation also gives an intuitive way to demonstrate the implication of non-trivial symmetry indicators: the transition between different forms of the localized vortex.

\subsection{Topological edge state between shifted bulks}
\begin{figure}[ht]
\centering
\includegraphics[width=12cm]{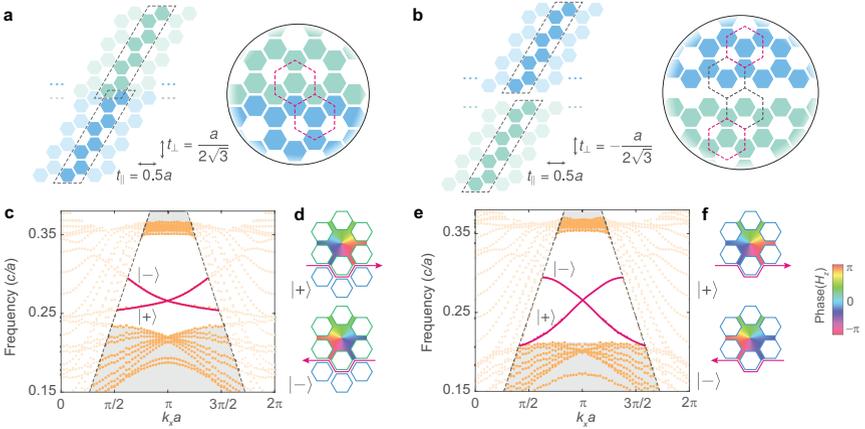}
\caption{Two types of defect line arrangements. \textbf{a} and \textbf{b} Illustration of the ``tight arrangement'' and the ``loose arrangement''. The structure on both sides of the line defect are identical and different colors are for indicating different topology. \textbf{c} and \textbf{e} Numerical solutions of the band diagram for tight and loose super-cell. Modes in the shadowed area are uninterested bulk/corner modes. \textbf{d} and \textbf{f} Edge state $H_z$ phase plot in a unit adjacent to the defect line. Purple lines show the position of the defect and the arrows indicate the direction of the energy flux.}\label{fig_edgemode}
\end{figure}

The edge states on the bulk interface induced by the topological phase difference is also protected by it \cite{Khalaf2018}. In this particular case, different topological phases can be achieved by merely shifting the bulks so that different boundaries are exposed. Fig. \ref{fig_edgemode}a and b illustrate two scenarios in this regard: both approaches involve a parallel shift of $a/2$ along the path, but one has a positive transverse shift pulling the bulks on the two sides close to each other (referred to as the ``tight arragnement'' below), and the other a negative transverse shift pulling two bulks away (referred to as the ``loose arrangment''). To tile these arrangements on the entire 2D plane with hexagonal units, one must use unit cells with the different symmetry indicators for different sides across the defect line. The numerical simulation results of the super-cell as in Fig. \ref{fig_edgemode}c-f verify the the existence of the topological edge modes for both scenarios. The two unidirectional modes are characterized by the opposite phase rotation direction within the unit cells adjacent to the defect line. 

\begin{figure}[ht]
\centering
\includegraphics[width=12cm]{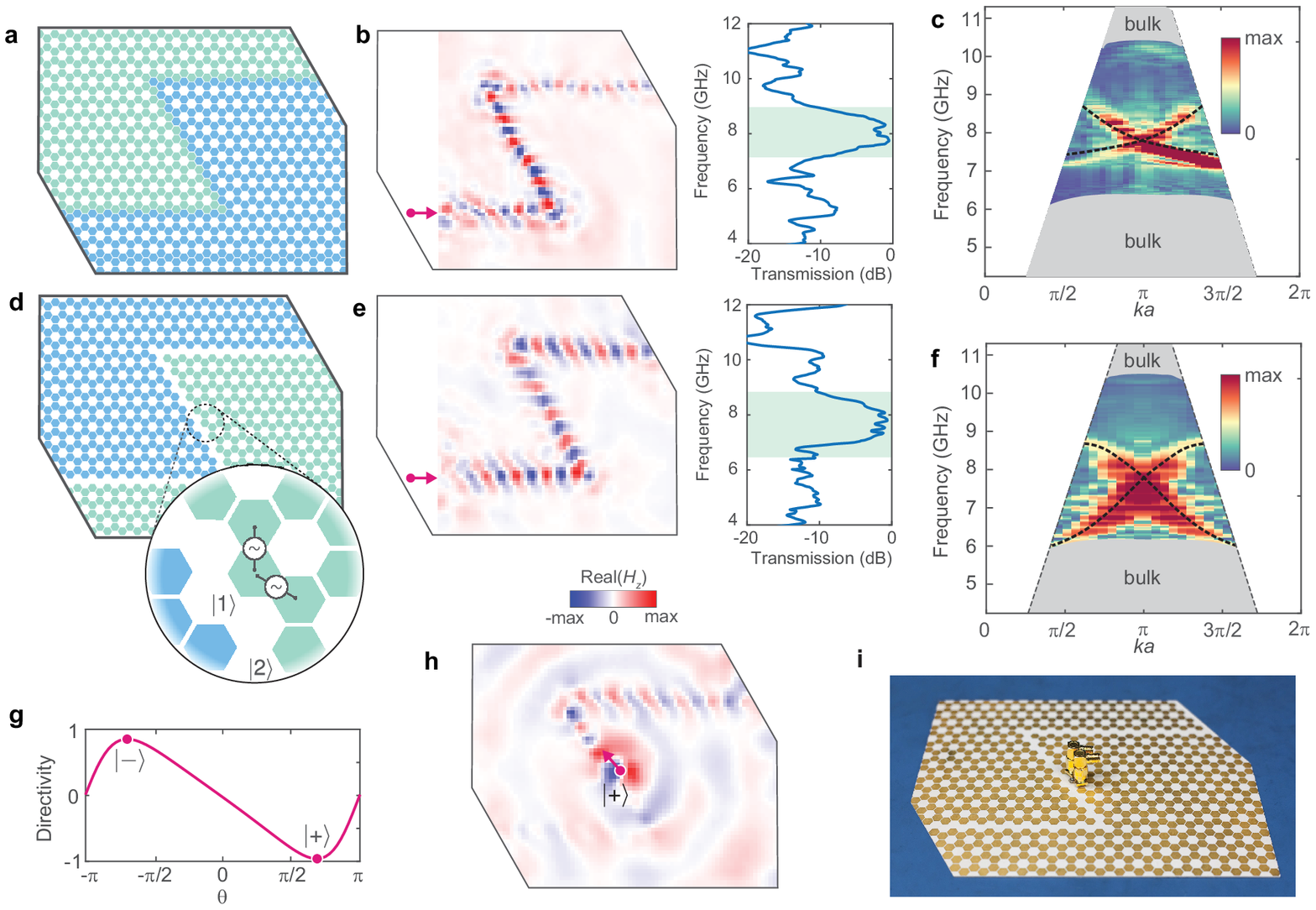}
\caption{The experiments verifying the predicted edge state. \textbf{a} and \textbf{d} Sketch of the PCB geometry implementing the tight and loose arrangement defect line with sharp turnings. \textbf{b} and \textbf{e} Results on the robustness against sharp corners of the tight arrangement and loose arrangement boards, respectively. Left insets: near-field $H_z$ scan results from the metallic side. The purple arrows represent the single excitation on one end of the waveguide. Right insets: transmission over frequency. Highlighted areas are the regions where edge modes locate on the band diagram. \textbf{c} and \textbf{f} LDOS plot for the tight and loose line defect, respectively. Dotted lines are simulated band structure. \textbf{g} Relationship between the reactivity and the phase delay between the two sources that excites the loose defect line. \textbf{h} Near-field $H_z$ plot, from the dielectric side of the loose defect line, with the paired excitation in the middle point of the waveguide, with a $2\pi/3$ phase delay. \textbf{i} The photo of the board, loose arrangement with the source pair.}\label{fig_expirement}
\end{figure}

In order to experimentally verify the topological modes for both tight and loose arrangement, we fabricate both defect line models with 120\textdegree~turns, as illustrated in Fig. \ref{fig_expirement}a and d, using a lattice constant $a=10.16$ mm. The energy can be well coupled into the desired surface wave mode by directly connecting the excitation RF port across two metallic patches. The transverse magnetic near-field scan can be then performed from either side of the board (See Methods, near-field scanning). The field intensity plots for both cases excited from one end of the waveguide, as shown in Fig.\ref{fig_expirement}b and e, exhibit constant amplitude along the defect line with no noticeable standing wave, indicating the robustness of the mode against the sharp turns. Furthermore, Fig.\ref{fig_expirement}c and f show the local density of states (LDOS) plot of the two arrangements by performing a Fourier transformation on the forward and backward mode $H_z$ along the center of the defect line path (See Methods, data processing). The results show good agreement with the numerical solution in Fig.\ref{fig_edgemode}.

Unlike many previous studies where special antennas are needed to excite the unidirectional mode, controlling the ``spin'' states are relatively easy for the model. For the loose arrangement, this extra degree of freedom can be obtained by implementing a pair of sources with equal amplitude and proper phase difference $\theta$, shown as Fig.\ref{fig_expirement}d and i. We build the source at the center of the waveguide, and the directivity can be then evaluated with:

\begin{equation}
    D=\frac{S_+-S_-}{S_++S_-}
\end{equation}
where $S_{+/-}$ are the energy flux toward the two directions, which can be represented by the squared amplitude of the field intensity. Fig. \ref{fig_expirement}g gives the variation of $D$ with respect to the phase difference. It can be found that $ \mid\pm\rangle = \mid1\rangle + e^{\mp i\frac{2\pi}{3}} \mid2\rangle$, in which $\mid1\rangle$ or $\mid2\rangle$ is the state each source excites on its own. This set of values agrees with the $\pm 2\pi$ phase accumulation per lap observed in Fig. \ref{fig_edgemode}f. A clear uni-directional edge state for the loose arrangement with the $ \mid+\rangle$ excitation is shown in Fig. \ref{fig_expirement}h. In a similar manner, the spin of the tight arrangement can be controlled but with three sources instead of two, with clockwise/counter-clockwise $\frac{2\pi}{3}$ phase progression (See Supplementary Note 2).

The fact that the designs presented above only adopt the translation operation endows it with advantages over conventional arrangements where various types of mirror symmetries are needed for cells on the different side of the waveguide. A straightforward benefit is greater reconfigurability: to modify the path of the waveguide, one only needs to apply proper slides on the lattice, instead of altering each unit cell itself.

\begin{figure}[ht]
\centering
\includegraphics[width=6cm]{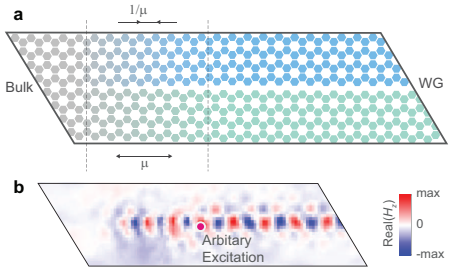}
\caption{Waveguide arrangement integrated with a termination. \textbf{a} The geometry. \textbf{b} Near-field $H_z$ profile, measured from the dielectric side, with a single excitation.}\label{fig_terminal}
\end{figure}

Another interesting property of this type of arrangement is the possibility to create a ``termination'' on one side of the path, as in Fig. \ref{fig_terminal}a. In between the dotted line regions, a gradual transverse translation and parallel shrink/expansion are applied to the lattice so that the waveguide gradually closes into the bulk. The parameter $\mu$ for the shrink/expansion is chosen to be very close to $1$ so that the band structure hardly changes. Consider an single excitation anywhere on the waveguide. The backward mode  $\mid-\rangle$ it excites is forced to convert to $\mid+\rangle$ near the termination point because of the existence of the bandgap in the whole patched region. As a result, the pure $\mid+\rangle$ mode can be obtained with a arbitrary excitation, as is verified experimentally in Fig. \ref{fig_terminal}b . It is worth noting that this terminating effect through gradual transform cannot be realized with conventional structures since the topological phase transition of the units requires a closed bandgap at some point, from which region the energy is able to leak out\cite{Wen2021}. 

\section{Discussion}
The realization of topological edge states in defect lines is demonstrated theoretically and is verified experimentally. Unlike earlier topological photonic designs, the edge mode is protected by the non-trivial symmetry indicators with real space gauge-dependency, rather than any topological invariant defined purely in reciprocal space. The design shows significant potential for practical application: it is ultra-thin and can be built with standard PCB techniques; the unidirectional states can be easily excited with a standard SMA port which make it easy to be integrated into conventional microwave systems; if printed on a flexible substrate, the design can be easily accommodated into a configurable version, with the potential to terminate the path at any point. Beyond these, the design is also a direct observation and verification of the new design principle which largely broadens the way to implement topological protection in photonic systems. The theory potentially provides a new perspective on some existing spin-related designs \cite{leFeber2015,Sollner2015} that were not examined from a topological point of view, and is also likely to inspire a new category of real-space induced topological photonic devices.

\section{Methods}
\subsection{Numerical simulation}
The numerical simulation results presented throughout the paper are obtained with the commercial FEM software Ansys HFSS. Eigenmode solvers are used for both unit-cell and super-cell simulations, solving 20 modes starting from $45$ GHz for a lattice constant $a=1$ mm. For unit-cell simulation, two types of setup are used: 1. rhombic unit-cell containing two PCB patches, with two pairs of periodic (primary-secondary) boundaries; 2. hexagonal unit-cells with three pairs of periodic boundaries. The boundary on the top and bottom are set to perfect H boundary, $1.2a$ away from the board. For the super-cell simulation, we select a parallelogram with a length of $8a$ in the transverse direction with one pair of periodic boundary conditions and four perfect H boundaries for the sides, top and bottom. 

\subsection{PCB Sample fabrication}
We used the parameter $a=10.16$ mm and $g_2=0.4064$ mm for the fabrication, so that the size of the edge state verification board is $203.2$ mm $\times130.6$ mm and the size for the termination board is $335.3$ mm$\times70.4$ mm. The boards are fabricated with standard PCB manufacturing techniques. The substrate being used is 1.27 mm (50 mil) Rogers RO3010 with 1 oz copper, ENIG finish, with no soldermask or silkscreen. For the excitation port, edge-launched L-shaped SMA connectors are modified into surface-mount configurations: the ground and center pins are cut down to 2 mm and soldered onto two metallic patches respectively.

\subsection{Experiment setup and data processing}
The near field scan are performed on an XY stage, with a resolution of $2$ mm$ \times2$ mm. We utilize a 2 cm thick foam block to support the sample, separating it from the metallic stage base. The vector network analyzer (VNA) Angilent E5071C is employed for the field measurement: port 1 is connected to the SMA port on the sample board and port 2 is connected to an $H_z$ probe 0.5 mm above the sample, which is made with a rigid SMA cable with a 5 mm loop at the end. The transmission between the ports $S_{21}$ from 4 GHz to 12 GHz si recorded. The measured transmission results are filtered with a 10 ns time gate in MATLAB to reduce the background noise: we perform the inverse Fourier transform on the frequency-dependent $S_{21}$, filter it with 20 ns rectangular gate around the peak and then convert it back to the frequency domain. For the transmission, we evaluate the relative value of $H_z^2$ between the end point of the third section of the waveguide and the start point of the first section in vicinity of the excitation. For the LDOS plot, we extract the one row of $H_{z}$ at the center of the defect line path and perform a Fourier transform, converting it to momentum space. This process is conducted for modes propagating in both directions (by excitation on different ends) and the results are added up to capture both bands. For uni-directional mode measurements, each port is excited separately with the other ports connected to broadband 50 $\Omega$ loads and the modes are superposed in MATLAB.


\bibliography{sn-bibliography}


\end{document}